\documentclass[11pt]{llncs}
\usepackage{graphicx}
\usepackage{cite}
\usepackage{url}
\usepackage{times}
\usepackage{mathptm}

\DeclareSymbolFont{AMSb}{U}{msb}{m}{n}
\DeclareSymbolFontAlphabet{\Bbb}{AMSb}

\def\Z{\ensuremath{\Bbb Z}}

\makeatletter
\def\hb@xt@{\hbox to }
\makeatother

\let\oldendproof\endproof
\def\endproof{\qed\oldendproof}

\def\Sc{\mathop{\rm Sc}}

\begin{document}

\title{Algorithms for Drawing Media} 

\author{David Eppstein\thanks{Supported in part by NSF grant CCR-9912338.}}

\institute{Computer Science Department\\
School of Information \& Computer Science\\
University of California, Irvine\\
\email{eppstein@uci.edu}}

\maketitle   

\begin{abstract}
We describe algorithms for drawing media, systems of states, tokens and actions that have state transition graphs in the form of partial cubes.  Our algorithms are based on two principles:
embedding the state transition graph in a low-dimensional integer lattice and projecting the lattice onto the plane, or drawing the medium as a
planar graph with centrally symmetric faces.
\end{abstract}

\section{Introduction}

Media~\cite{cs.DS/0206033,FalOvc-DAM-02} are systems of states, tokens,
and actions of tokens on states that arise in political choice theory and that can also be used to represent many familiar geometric and combinatorial systems such as hyperplane arrangements, permutations, partial orders, and phylogenetic trees.
In view of their importance in modeling social and combinatorial systems, we would like to have efficient algorithms for drawing media as state-transition graphs in a way that makes the action of each token apparent.  In this paper we describe several such algorithms.

Formally, a {\em medium} consists of a finite set of {\em states} transformed by the actions
of a set of {\em tokens}.  The concatenation of tokens is called a {\em message}; we use upper case letters to denote states, and lower case letters to denote tokens and messages, so $Sw$ denotes the state formed by applying the tokens in message $w$ to state $S$.
A token $t$ is said to be {\em effective} for $S$ if $St\neq S$, and a message $w$ is {\em stepwise effective} for $S$ if each successive token in the sequence of transformations of $S$ by $w$ is effective. 
For a set of states and tokens to form a medium, it must satisfy the following axioms:

\begin{enumerate}
\item For each token $t$ there is a unique {\em reverse} token $\tilde t$ such that, for any two states $S\neq Q$, $St=Q$ iff $Q\tilde t=S$.
\item For any two distinct states $S$, $Q$, there exists a message $w$ with $Sw=Q$
such that $w$ does not contain both $t$ and $\tilde t$ for any token $t$
(we say that a message that does not contain the reverse of any of its tokens is {\em consistent}).
\item  If message $w$  is stepwise effective for $S$, then  $Sw=S$ if and only if the number of copies of $t$ in $w$ equals the number of copies of $\tilde t$ for each token $t$.
\item If $Sw=Qz$, $w$ is stepwise effective for $S$, $z$ is stepwise effective for $Q$, and
both $w$ and $z$ are consistent, then $wz$ is consistent.
\end{enumerate}

The states and state transitions of a medium can also be viewed as a graph, and the axioms defining media imply that these graphs are {\em partial cubes}; a partial cube~\cite{ImrKla-00} is a graph with vertices that can be mapped to vertices of a hypercube $\{0,1\}^d$ in such a way that the graph distance equals the $L_1$ distance in the hypercube. For media, we can find such a mapping by choosing arbitrarily a state $S$ in the medium, and assigning any state $S'$ a coordinate per token $t$ that is $1$ when a consistent path from $S$ to $S'$ contains $t$ and $0$ otherwise. Conversely, any $d$-dimensional partial cube gives rise to a medium with its vertices as states and with $2d$ tokens; the action of any token is to change one of the partial cube coordinates to a zero or to a one, if it does not already have that value and if such a change would produce the coordinates of another vertex of the partial cube.

We assume throughout, as in~\cite{cs.DS/0206033}, that we are given as input an explicit description of the states, tokens, and actions of a medium.  However, our algorithms are equally applicable to any partial cube or family of partial cube graphs such as the median graphs.  If a partial cube representation is not already known, such a representation can be found (and the corresponding medium constructed) in time $O(mn)$ via known algorithms~\cite{AurHag-MST-95,ImrKla-EJC-96,Win-DAM-84,ImrKla-00}.

\section{Lattice dimension}

\begin{figure}[t]
\centering
\includegraphics[width=5in]{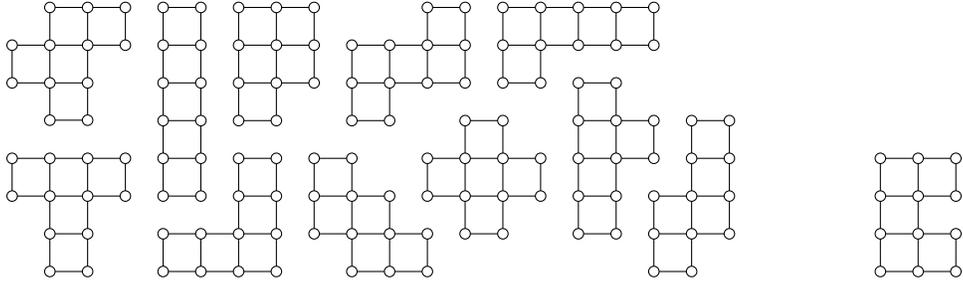}
\caption{11 of the 12 pentominos represent isometric lattice embeddings of media.  The twelfth, the U pentomino, does not, because a pair of vertices that are three edges apart in the graph have placements that are only one unit apart.}
\label{fig:pentomino}
\end{figure}

As we have seen, media can be embedded {\em isometrically} (that is, in a distance-preserving way) into hypercubes $\{0,1\}^d$ (with $L_1$ distance), and hypercubes can be embedded isometrically into integer lattices $\Z^d$, so by transitivity media can be embedded isometrically onto integer lattices.  Conversely any finite isometric subset of an integer lattice forms a partial cube and corresponds as described above to a medium.

If the dimension of the lattice in which a medium is embedded is low, we may be able to use the embedding as part of an effective drawing algorithm.
For instance, if a medium $M$ can be embedded isometrically onto the planar integer lattice
$\Z^2$, then we can use the lattice positions as vertex coordinates of a drawing in which each edge is a vertical or horizontal unit segment (Figure~\ref{fig:pentomino}).  If $M$ can be embedded isometrically onto the cubic lattice $\Z^3$, in such a way that the projection onto a plane perpendicular to the vector $(1,1,1)$ projects different vertices to distinct positions in the plane, then this projection produces a planar graph drawing in which the edges are unit vectors at $60^\circ$ and $120^\circ$ angles
(Figure~\ref{fig:order}, center; the right drawing in the same figure could have been produced in this way but was actually produced by a different algorithm.)

Recently, we showed that the {\em lattice dimension} of a medium or partial cube, that is, the minimum dimension of a lattice $\Z^d$ into which it may be isometrically embedded, may be determined in polynomial time~\cite{Epp-EJC-?}.  We now briefly our algorithm for finding low-dimensional lattice embeddings.

\begin{figure}[t]
\centering
\includegraphics[width=3.5in]{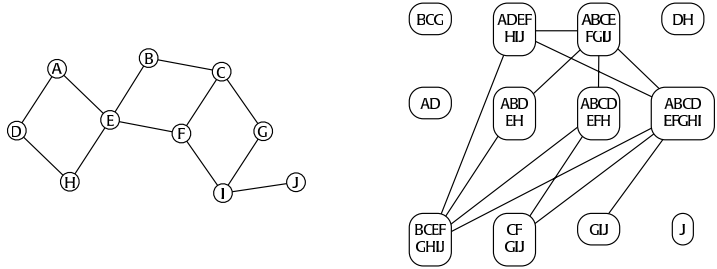}
\caption{A medium, left, and its semicube graph, right. From~\cite{Epp-EJC-?}.}
\label{fig:semicube}
\end{figure}

\begin{figure}[t]
\centering
\includegraphics[width=3.5in]{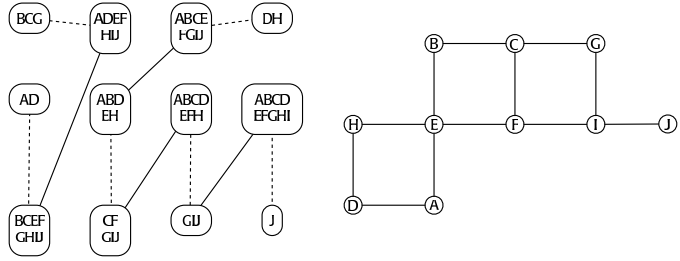}
\caption{A matching in the semicube graph (left, solid edges) completed to a set of paths by adding edges from each semicube to its complement (left, dashed edges), and the corresponding lattice embedding of the original medium (right). From~\cite{Epp-EJC-?}.}
\label{fig:matched}
\end{figure}

Suppose
we are given an undirected graph $G$ and an isometry $\mu:G\mapsto\{0,1\}^\tau$ from $G$ to the hypercube $\{0,1\}^\tau$ of dimension $\tau$.
Let $\mu_i:G\mapsto \{0,1\}$ map each vertex $v$ of $G$ to the $i$th coordinate of $\mu(v)$,
and assume that each coordinate $\mu_i$ takes on both value $0$ and $1$ for at least one point .
From $G$ and $\mu$ we can define $2\tau$ distinct {\em semicubes}
$S_{i,\chi}=\{v\in V(G)\mid \mu_i(v)=\chi\}$, for any pair $i,\chi$ with
 $0\le i<\tau$ and $\chi\in\{0,1\}$.
We now construct a new graph  $\Sc(G)$, which we call the {\em semicube graph} of $G$.
We include in $\Sc(G)$ a set of $2\tau$ vertices $u_{i,\chi}$, $0\le i<\tau$ and $\chi\in\{0,1\}$.
We include an edge in $\Sc(G)$ between $u_{a,b}$ and $u_{c,d}$
whenever $S_{a,b}\cup S_{c,d}=V(G)$ and $S_{a,b}\cap S_{c,d}\ne\emptyset$; that is, whenever the corresponding two
semicubes cover all the vertices of $G$ non-disjointly.
Although defined from some particular isometry $\mu$, the semicube graph turns out to be independent of the choice of~$\mu$.
An example of a partial cube $G$ and its semicube graph $\Sc(G)$ is shown in Figure~\ref{fig:semicube}.  The main result of~\cite{Epp-EJC-?} is that the lattice dimension of $G$ can
be determined from the cardinality of a maximum matching in $\Sc(G)$:

\begin{theorem}[Eppstein~\cite{Epp-EJC-?}]
If $G$ is a partial cube with isometric dimension $\tau$, then the lattice dimension of $G$ is
$d=\tau-|M|$ where $M$ is any maximum matching in $\Sc(G)$.
\end{theorem}

More specifically, we can extend a matching in $\Sc(G)$ to a collection of $d$ paths by adding to the matching an edge from each semicube to its complement.  The $d$th coordinate of a vertex in the lattice embedding equals the number of semicubes that contain the vertex in even positions along the $d$th path.

We can use this result as part of a graph drawing system, by embedding our input medium in the lattice of the lowest possible dimension and then projecting that lattice onto the plane.  For two-dimensional lattices, no projection is needed, and we have already discussed projection of certain three-dimensional integer lattices onto two-dimensional triangular lattices.  We discuss more general techniques for lattice projection in the next section.
We note that, in order to find the minimum dimension lattice embedding, it is essential that we require the embedding to be isometric.  Even for trees (a very special case of partial cubes) it is NP-complete to find an embedding into $\Z^2$ with unit length edges that is not required to be distance-preserving~\cite{BhaCos-IPL-87}.  However a tree embeds isometrically in $\Z^2$ if and only if it has at most four leaves~\cite{math.CO/0402246}.

\section{Drawing high-dimensional lattice graphs}

We have seen that two-dimensional lattice embeddings of media, and some three-dimensional lattice embeddings, give rise to planar graph drawings with all edges short and well separated by angles.
However, we are also interested in drawing media that may not have low dimensional embeddings.  We describe here a method for finding drawings with the following properties:

\begin{enumerate}
\item\label{prop:distinct-vertices}
All vertices are assigned distinct integer coordinates in $\Z^2$.
\item\label{prop:straight-edges}
All edges are drawn as straight line segments.
\item\label{prop:vertex-edge-separation}
No edge passes closer than unit distance to a vertex that is not one of its endpoints.
\item\label{prop:translate-if-parallel}
The line segments representing two edges of the drawing are translates
of each other if and only if the two edges are parallel in the lattice embedding.
\item\label{prop:product-area}
The medium corresponding to any Cartesian product of intervals
$[a_0,b_0]\times[a_1,b_1]\times\cdots[a_{d-1},b_{d-1}]$ is drawn in 
area $O(n^2)$, where $n$ denotes the number of states of the medium.
\end{enumerate}

Because of property~\ref{prop:translate-if-parallel}, the lattice embedding and hence the medium structure of the state transition graph can be read from the drawing.  To achieve these, properties,
we map $\Z^d$ to $\Z^2$ linearly, by choosing wo vectors $X$ and $Y\in\Z^d$, and mapping any point $p\in\Z^d$ to the point $(X\cdot p,Y\cdot p)\in\Z^2$.  We now describe how these vectors $X$ and $Y$ are chosen.

If $L\subset\Z^d$ is the set of vertex placements in the lattice embedding of our input medium,
define a {\em slice} $L_{i,j}=\{p\in L\mid p_i=j\}$ to be the subset of vertices having $i$th coordinate equal to~$j$.  We choose the coordinates $X_i$ sequentially, from smaller $i$ to larger, so that all slices $L_{i,j}$ are separated from each other in the range of $x$-coordinates they are placed in.
Specifically, set $X_0=0$.  Then, for $i>0$, define
$$
X_i = \max_j
(\min_{p\in L_{i,j}} \sum_{k=0}^{i-1} X_k p_k -
\max_{q\in L_{i,j-1}} \sum_{k=0}^{i-1} X_k q_k),
$$
where the outer maximization is over all $j$ such that
$L_{i,j}$ and $L_{i,j-1}$ are both nonempty.
We define $Y$ similarly, but we choose its coordinates in the opposite order, from larger $i$ to smaller:
$Y_{d-1}=0$, and
$$
Y_i=\max_j
(\min_{p\in L_{i,j}} \sum_{k=i+1}^{d-1} X_k p_k -
\max_{q\in L_{i,j-1}} \sum_{k=i+1}^{d-1} X_k q_k).
$$

\begin{figure}[t]
\centering
\includegraphics[scale=0.25]{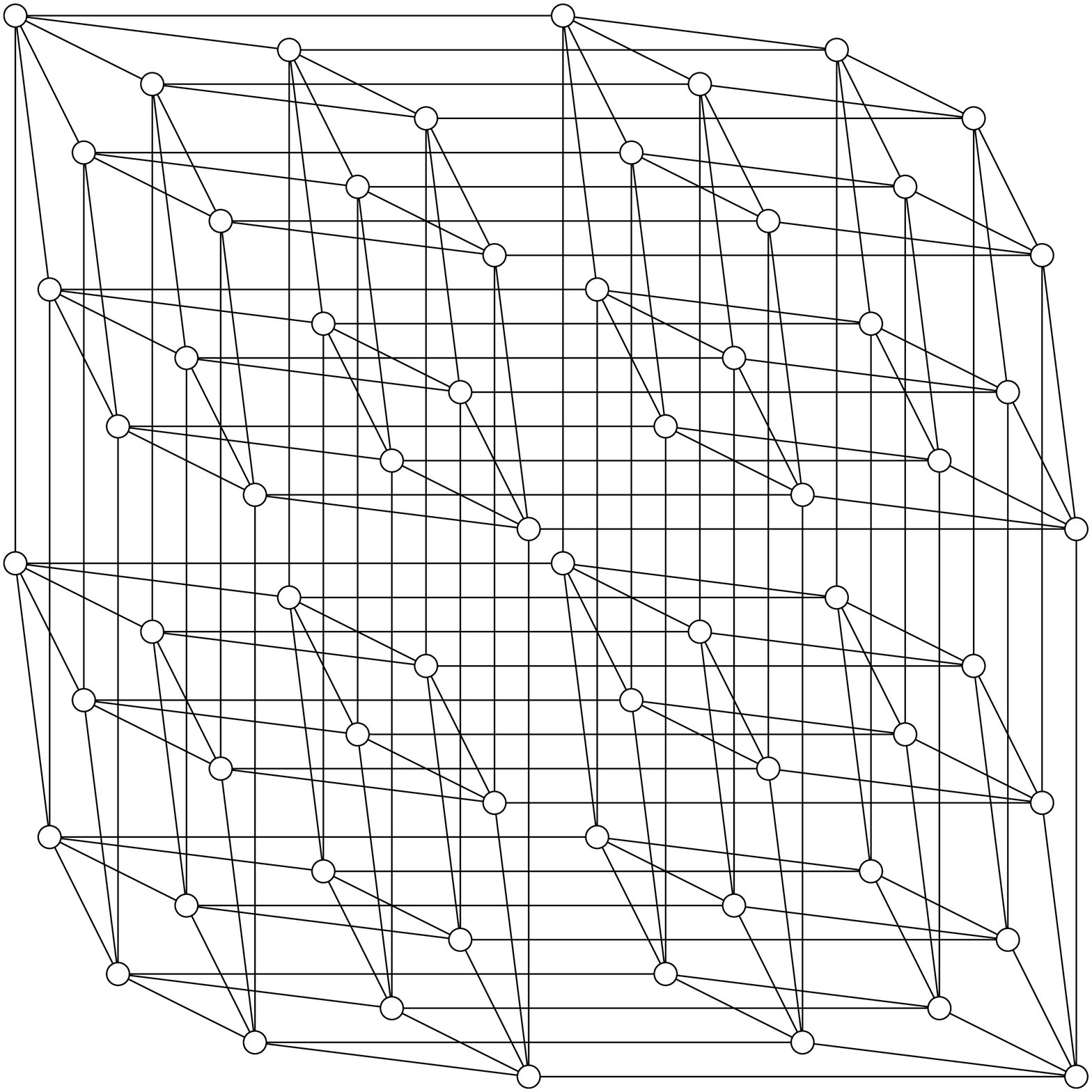}\qquad
\includegraphics[scale=0.25]{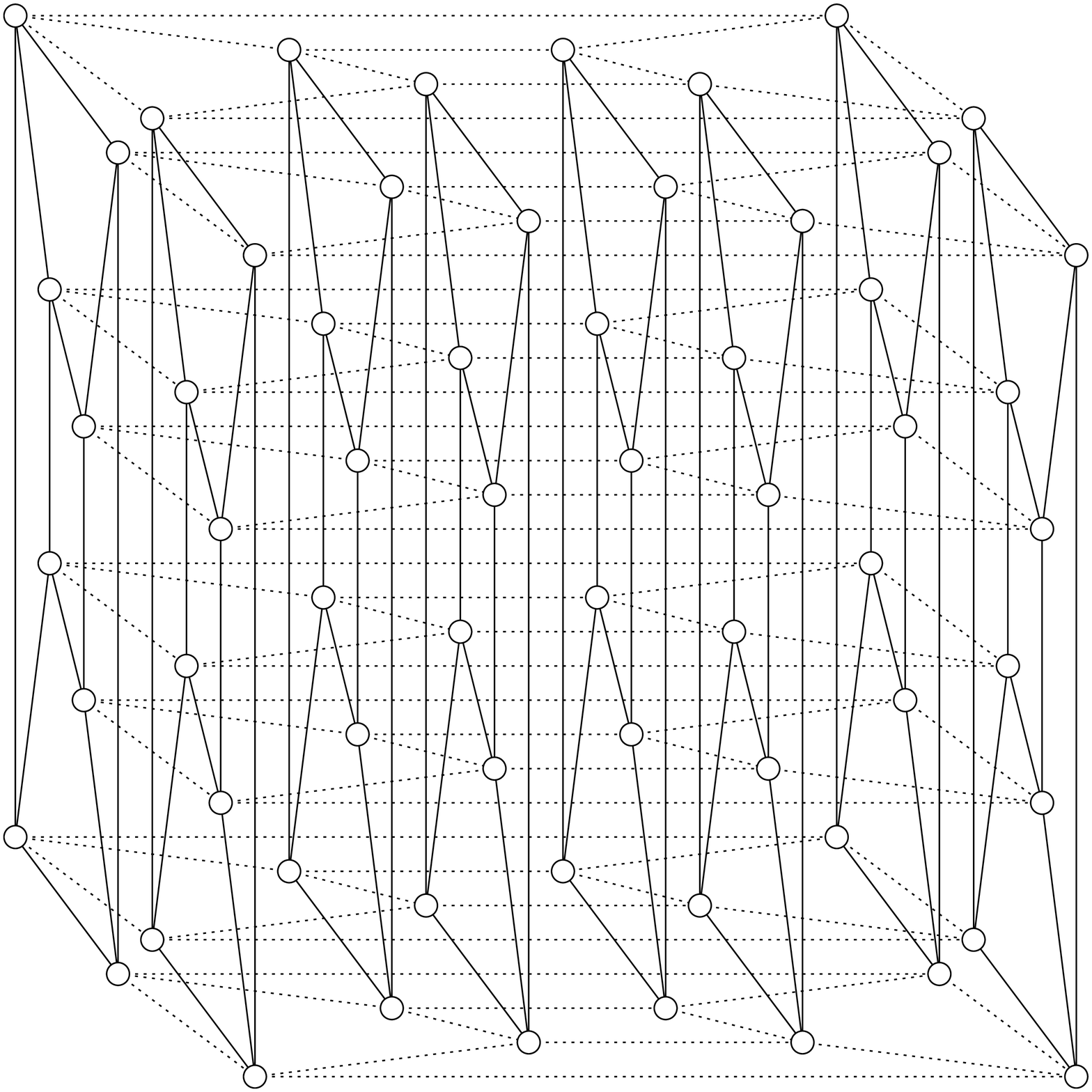}
\caption{Left: lattice drawing of six-dimensional hypercube;
right: a drawing with geometric thickness two is possible, but the vertex placement is less regular
and edges formed by actions of the same token are not all drawn parallel.}
\label{fig:cube6}
\end{figure}

\begin{theorem}
The projection method described above satisfies the properties~\ref{prop:distinct-vertices}--\ref{prop:product-area} enumerated above.
\end{theorem}

\begin{proof}
Property~\ref{prop:straight-edges} and property~\ref{prop:translate-if-parallel}
follow immediately from the fact that we our drawing is formed by projecting $\Z^d$ linearly onto $\Z^2$, and from the fact that the formulas used to calculate $X$ and $Y$ assign different values to different coordinates of these vectors.

It is clear from this construction that all vertices are assigned distinct coordinates (property~\ref{prop:distinct-vertices}):
for, if vertices $p$ and $q$ differ in the $i$th coordinates of their lattice embeddings, they belong to different slices $L_{i,j}$ and $L_{i,j'}$ and are assigned $X$ coordinates that differ by
at least $X_i$ (unless $i=X_i=0$ in which case their $Y$ coordinates differ by at least $Y_i$).

The separation between vertices and edges (property~\ref{prop:vertex-edge-separation}) is almost equally easy to verify: consider the case of three vertices $p$, $q$, and $r$, with an edge $pq$ to be separated from $r$.  Since $p$ and $q$ are connected by an edge, their lattice embeddings must differ in only a single coordinate~$i$.
If $r$ differs from $p$ and $q$ only in the same coordinate, it is separated from edge $pq$ by
a multiple of $(X_i,Y_i)$.  Otherwise, there is some coordinate $i'\neq i$ in which $r$ differs
from both $p$ and $q$.  If $i'>i$, the construction ensures that the slice $L_{i',j}$ containing $pq$ is well separated in the $x$-coordinate from the slice $L_{i',j'}$ containing $r$, and if $i'<i$ these slices are well separated in the $y$ coordinate.

Finally, we consider property~\ref{prop:product-area}.
For Cartesian products of intervals, in the formula for $X_i$, the value for the subexpression
$\min_{p\in L_{i,j}} \sum_{k=0}^{i-1} X_k p_k$ is the same for all $j$ considered in the outer maximization, and the value for the subexpression
$\max_{q\in L_{i,j-1}} \sum_{k=0}^{i-1} X_k q_k$ is also the same for all $j$ considered in the outer maximization, because the slices are all just translates of each other.
Therefore, there is no gap in $x$-coordinates between vertex placements of each successive slice of the medium.  Since our drawings of these media have vertices occupying contiguous integer $x$ coordinates and (by a symmetric argument) $y$ coordinates, the total area is at most $n^2$.
\end{proof}

When applied to a hypercube, the coordinates $X_i$ become powers of two, and this vertex placement algorithm produces a uniform placement of vertices (Figure~\ref{fig:cube6}, left) closely related to the Hammersley point set commonly used in numerical computation and computer graphics for its low discrepancy properties~\cite{WonLukHen-JGT-97}.  Other examples of drawings produced by this method can be seen in
Figures~\ref{fig:nosfp}, \ref{fig:grids3d}, and \ref{fig:order}(left).

\section{Symmetric planar drawings}

\begin{figure}[t]
\centering
\includegraphics[width=5in]{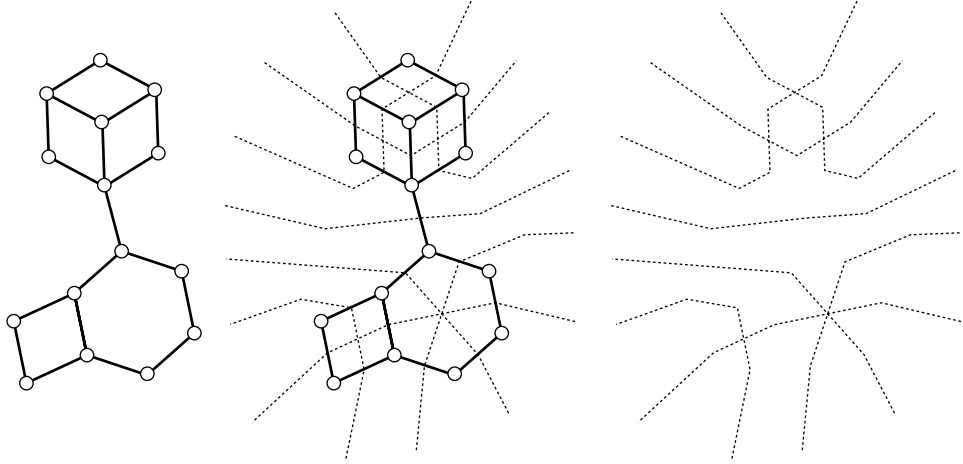}
\caption{Left: a graph with a symmetric planar drawing; center: connecting opposite pairs of edge midpoints produces a weak pseudoline arrangement; right: the arrangement.}
\label{fig:wpla}
\end{figure}

Our two-dimensional and projected three-dimensional lattice drawings are planar
(no two edges cross) and each internal face is symmetric (squares for two-dimensional lattices,
60$^\circ$-120$^\circ$ rhombi and regular hexagons for projected three-dimensional lattices).
We now describe a different type of drawing of the state-transition graphs of media
as planar graphs, generalizing this symmetry property.  Specifically, we seek straight-line planar drawings in which each internal face is strictly convex and centrally symmetric; we call such a drawing
a {\em symmetric planar drawing}.

A {\em weak arrangement of pseudolines}~\cite{FraOss-GD-03} is a collection of curves
in the plane, each homeomorphic to a line, such that any pair of curves in the collection has at most one point of intersection, and such that if any two curves intersect then they cross properly at their intersection point.  Weak arrangements of pseudolines generalize pseudoline arrangements~\cite{GooPol-NTDCG-93}
and hyperbolic line arrangements, and are a special case of the {\em extendible pseudosegment arrangements} defined by Chan~\cite{Cha-DCG-03}.  Any weak pseudoline arrangement with $n$ pseudolines partitions the plane into at least $n+1$ and at most $n(n+1)/2+1$ {\em cells},
connected components of the set of points that do not belong to any pseudoline.
Any pseudoline in the arrangement can be partitioned into {\em nodes} (crossing points) and {\em arcs} (connected components of the complement of the crossing points); we use this terminology to avoid confusion with the vertices and edges of the medium state-transition graphs we hope to draw.
Each arc is adjacent to two cells and two nodes.
We define the {\em dual} of a weak pseudoline arrangement to be the graph having a vertex for each
cell of the arrangement and an edge connecting the vertices dual to any two cells that share a common arc; this duality places the graph's vertices in one-to-one correspondence with the arrangement's cells, and the graph's edges in one-to-one correspondence with the arrangement's arcs.

\begin{lemma}
\label{lem:spd2wpla}
If $G$ has a symmetric planar drawing, then $G$ is the dual of a weak pseudoline arrangement.
\end{lemma}

\begin{proof}
Draw a collection of line segments connecting the opposite pairs of edge midpoints in each interior face of the drawing.  In the exterior face, continue the segments from each midpoint into curves extending to infinity without any additional crossings.  An example depicting this collection of line segments and continuation curves is shown in Figure~\ref{fig:wpla}.  Each curve in the resulting arrangement is a pseudoline that passes through a collection of line segments in the drawing of $G$ that are translates of each other. If we arbitrarily orient each pseudoline, then any crossing of two pseudolines $\ell_1$ and $\ell_2$ is either clockwise (the outward direction along $\ell_2$ is clockwise of the outward direction along $\ell_1$, with respect to the two inward directions) or counterclockwise, and the orientation of the crossing
can be determined by the relative slopes of the segments crossed by $\ell_1$ and $\ell_2$ respectively.
Therefore, all crossings between the same pair of pseudolines have the same orientation.  But if two oriented pseudolines cross multiple times, the crossings must alternate orientations, so any two pseudolines cross at most once and we have a weak pseudoline arrangement.  Each edge of $G$ crosses a single arc of the arrangement.  Each pseudoline of the arrangement has one node per face of $G$ it passes through, so each arc of the arrangement is crossed by a single edge of $G$.
Therefore, $G$ is the dual of the weak pseudoline arrangement we have constructed.
\end{proof}

\begin{lemma}
\label{lem:wpla-medium}
If $G$ is the dual of a weak pseudoline arrangement, then $G$ is the state transition graph of a medium.
\end{lemma}

\begin{proof}
Each pseudoline $\ell_i$ partitions the plane into two regions which we call $\ell_i^+$ and $\ell_i^-$,
choosing arbitrarily which of the two regions is given which name.  We then construct a system of states and tokens, in which we have one state per cell of the arrangement (equivalently, per vertex of $G$)
and one token per region $\ell_i^\pm$.  The action $St$ is defined to transform state $S$ into state $Q$,
if $Q$ is in region $t$ and can be reached from $S$ by crossing a single arc of the arrangement;
otherwise $St=S$.  It is easy to verify that this system satisfies the axioms of a medium and has $G$ as its state transition graph.
\end{proof}

\begin{figure}[t]
\centering
\includegraphics[height=1in]{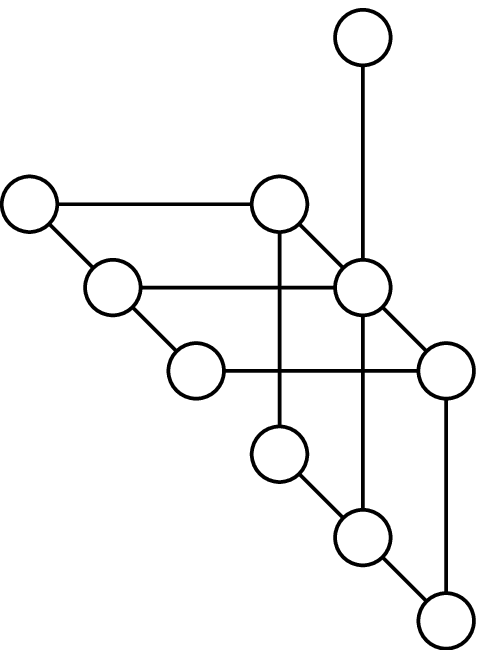}\qquad
\includegraphics[height=0.5in]{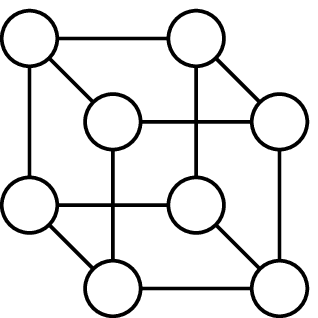}\qquad
\includegraphics[height=0.5in]{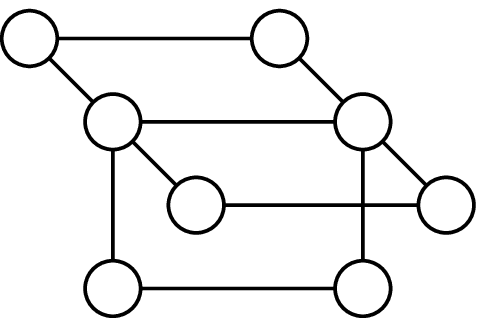}\qquad
\includegraphics[height=0.875in]{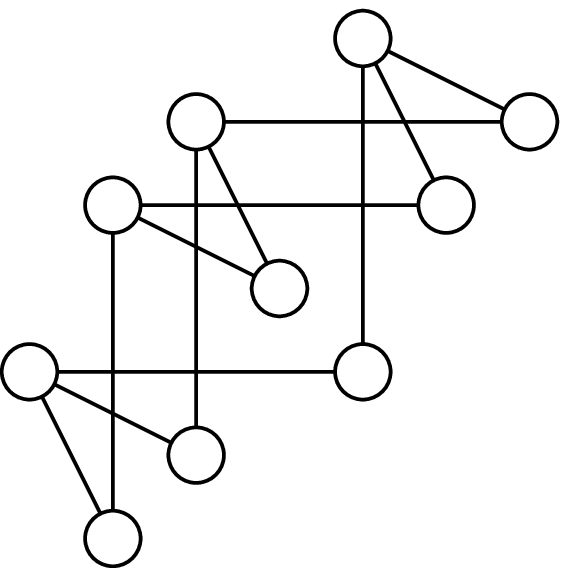}
\caption{Media with planar state-transition graphs but with no symmetric planar drawing.}
\label{fig:nosfp}
\end{figure}

By these lemmas, every symmetric planar drawing represents the state transition graph of a medium.  However, not every medium, and not even every medium with a planar state transition graph, has such a drawing; see for instance Figure~\ref{fig:nosfp}, the medium in Figure~\ref{fig:grids3d}(right), and the permutahedron in Figure~\ref{fig:order}(left) for media that have planar state transition graphs but no symmetric planar drawing.

\begin{figure}[t]
\centering
\includegraphics[width=4in]{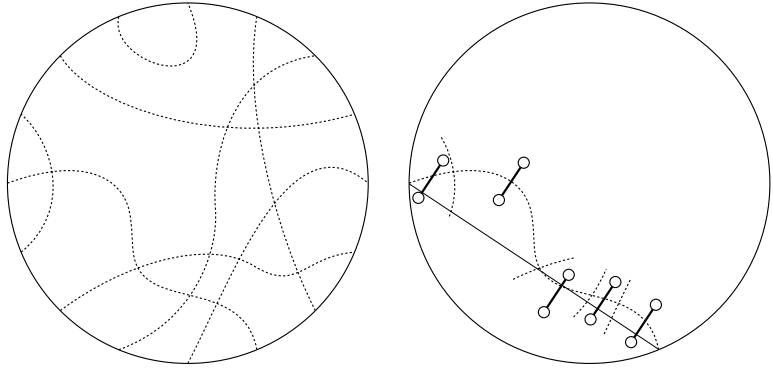}
\caption{Converting a weak pseudoline arrangement into a symmetric planar drawing.  Left: arrangement drawn inside a circle $O$ such that crossings with $O$ are equally spaced around the circle.  Right: edges dual to arcs of $\ell_i$ are drawn as unit length and perpendicular to the chord through the points where $\ell_i$ crosses $O$.}
\label{fig:wpla2spf}
\end{figure}

\begin{figure}[t]
\centering
\includegraphics[width=4in]{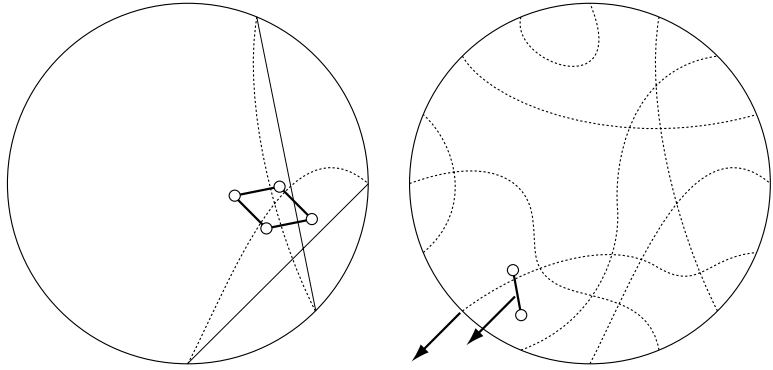}
\caption{Steps in planarity proof for construction of symmetric planar drawing from
weak pseudoline arrangement.  Left: angles between edges around internal faces equal the angles between corresponding pairs of chords.  Right: any point of an external face edge is the starting point of a ray not crossed by any other external face edge, formed by translating a ray perpendicular to $O$ from the point
where the arc dual to the edge crosses~$O$.}
\label{fig:spf-is-planar}
\end{figure}

\begin{lemma}
\label{lem:wpla2spd}
If $G$ is the dual of a weak pseudoline arrangement, then $G$ has a symmetric planar drawing.
\end{lemma}

\begin{proof}
Suppose $G$ is dual to a weak pseudoline arrangement $\cal A$; the duality fixes a choice of planar
embedding of $G$ as well as determining which faces of that embedding are internal and external.
We denote by $|\cal A|$ the number of pseudolines in $\cal A$.
Let $O$ be a circle (the size and placement of the circle within the plane being irrelevant to our construction), and deform $\cal A$ as necessary so that each pseudoline crosses $O$, with all nodes interior to $O$, and so that the $2|\cal A|$ points where pseudolines cross $O$ are spaced at equal distances around the perimeter of $O$ (Figure~\ref{fig:wpla2spf}, left).
Then, for each pseudoline $\ell_i$ of $\cal A$, let $c_i$ be the chord of $O$ connecting the two points where $\ell_i$ crosses $O$.  We will draw $G$ in such a way that the edges of $G$ that are dual to arcs of~$\ell_i$ are drawn as unit length segments perpendicular to $c_i$ (Figure~\ref{fig:wpla2spf}, right).
To do so, choose an arbitrary starting vertex $v_0$ of $G$, and place it arbitrarily within the plane.
Then, the placement of any other vertex $v_i$ of $G$ can be found by following a path from $v_0$ to $v_i$ in $G$, and for each edge of the path moving unit distance (starting from the location of $v_0$) in the direction determined for that edge as described above, placing $v_i$ at the point reached by this motion when the end of the path is reached.  It is straightforward to show from Lemma~\ref{lem:wpla-medium} and the axioms defining a medium that this vertex placement does not depend on the choice of the path from $v_0$ to $v_i$, and that if all vertices are placed in this way then all edges of $G$ will be unit length and perpendicular to their corresponding chords~$c_i$.
Thus, we have a drawing of $G$, in which we can identify sets of edges corresponding to the faces of $G$.  It remains to show that this drawing is symmetric planar; that is, that the faces are convex polygons and are connected to each other in a non-crossing way.

If pseudolines $\ell_i$ and $\ell_j$ meet in a node of $\cal A$, that node is dual to an internal face $f$ of $G$.  Suppose further that, in $G$, the pairs of edges of $f$ that are dual to the arcs of $\ell_i$ and $\ell_j$ on either side of the node meet in two vertices of $f$.
Then the exterior angles of $f$ at these vertices are, by construction, equal to the angle
between chords $c_i$ and $c_j$, which must also cross (Figure~\ref{fig:spf-is-planar}, left).  It follows from this that $f$ is drawn with all vertices convex and with total exterior angle (winding number) $360^\circ$, so $f$ is drawn as a correctly oriented convex polygon.  A similar argument shows that the angles of the adjacent pairs of edges around each vertex (where adjacency is determined according to the planar embedding dual to $\cal A$) and the total exterior angle of the external face of $G$ are $360^\circ$, so our drawing is locally planar.  However, we still must verify that the external face of $G$ is drawn in a non-self-crossing way, as that does not follow from the above assertions.

To verify this, we prove a stronger property: that, for each point $p$ on the external face $e$ of $G$, there is a ray $r_p$ with $p$ as its starting point that extends to infinity without crossing any other edge or vertex of $G$. If $p$ belongs to a bridge or is an articulation point of $G$, there are two such rays, one on each side of $p$.  Therefore, $e$ must be a simple polygon without self-crossings, and our drawing of $G$ must be a symmetric planar drawing.
To find $r_p$, suppose that $p$ belongs to an edge dual to an arc of pseudoline $\ell_i$.
That arc must contain an endpoint of $c_i$ (else the edge would not belong to $e$);
let $x$ be that endpoint.  Let $r_x$ be a ray starting from $x$, perpendicular to and disjoint from $O$, and let $r_p$ be formed by translating $r_x$ to $p$ (Figure~\ref{fig:spf-is-planar}, right).
Then an edge of the outer face of $f$ (if such an edge existed) dual to an arc of pseudoline $\ell_j$,
where the two endpoints of $c_j$ are near to and both clockwise of $x$, would have a slope slightly clockwise to that of $r_p$; similarly if the two endpoints of $c_j$ are near to and both counterclockwise of $x$ the slope is slightly counterclockwise to that of $r_p$; all other edges have angles between these two extremes, and in no case does any edge of $f$ have a slope in the appropriate range to block $r_p$.
\end{proof}

\begin{lemma}
\label{lem:2c-unique-embed}
If $G$ is biconnected, there is at most one planar embedding of $G$ that is dual to a weak pseudoline arrangement, and that embedding (if it exists) can be found in time $O(n)$.
\end{lemma}

\begin{proof}
We use a standard technique in graph drawing and planar embedding problems, the SPQR tree~\cite{DiBTam-FOCS-89,Mut-ICALP-03}, which can be used to represent the triconnected components of a graph.  Each node $v$ in the SPQR tree of $G$ has associated with it a multigraph $G_v$
consisting of some subset of vertices of $G$, edges of $G$, and {\em virtual edges} representing contracted parts of the remaining graph that can be separated from the edges of $G_v$ by a {\em split pair} of vertices (the endpoints of the virtual edge).  The non-virtual edges of $G$ are partitioned in this way among the nodes of the SPQR tree.  If two nodes are connected by an edge in the SPQR tree, then each has a virtual edge connecting two vertices that are shared between the two nodes.
We root the SPQR tree arbitrarily;
let $(s_v,t_v)$ denote the split pair connecting a non-root node $v$ to its parent,
and let $H_v$ denote the graph (with one virtual edge) represented by the SPQR subtree rooted at~$v$.
We work bottom up in the rooted tree, showing by induction on tree size that the following properties
hold for each node of the tree:
\begin{enumerate}
\item Each graph $H_v$ has at most one planar embedding that can be part of an embedding of $G$ dual to a weak pseudoline arrangement.
\item If $v$ is a non-root node, and $G$ is dual to a weak pseudoline arrangement, then edge $s_vt_v$ belongs
to the outer face of the embedding of~$H_v$.
\item If $v$ is a non-root node, form the path $p_v$ by removing virtual edge
$s_vt_v$ from the outer face of  $H_v$.  Then $p_v$ must lie along the outer face of any embedding of $G$ dual to a weak pseudoline arrangement.
\end{enumerate}
SPQR trees are divided into four different cases (represented by the initials S, P, Q, and R) and our proof follows the same case analysis:

\begin{description}
\itemsep\smallskipamount
\item[Trivial case:]
If $G_v$ consists of a single graph edge and a single virtual edge (a Q-node), then clearly
there can only be one planar embedding (up to reflection) of $G_v$.

\item[Parallel case:]
If $G_v$ consists of three or more edges connecting $(s_v,t_v)$ (a P-node), at most one of these edges can be non-virtual.  In this case, $G$ can only be dual to a weak pseudoline arrangement (and therefore have a symmetric planar drawing) if $G_v$ has three edges, one of which is non-virtual.
For, in any other case, a drawing of $G$ would have a nontrivial split graph of $(s_v,t_v)$
sandwiched between two interior faces bounded by two other such split graphs, and one or both of these two interior faces could not be drawn strictly convexly.
Thus, we assume $G_v$ has this special form.  If $v$ is the root of the SPQR tree, it has two children $u$ and $w$.  In this case, the embedding of $H_v=G$ must be formed by placing $H_u$ and $H_w$ on opposite sides of the edge $s_vt_v$,
with the paths $p_u$ and $p_v$ facing outwards.   If these conditions are satisfied, we have found as desired a unique embedding for~$G$.  If $v$ is not the root, it has one child $u$, and $H_v$ differs from $H_u$ by the addition of a single non-virtual edge $s_vt_v$.  As before, the non-virtual edge must be sandwiched between the two other parts of $G$, so the only possible embedding of $H_v$ is to place the non-virtual edge $s_vt_v$ parallel to
the virtual edge of $H_u$ connecting the same two vertices, on the internal side of this virtual edge.

\item[Series case:]
If $G_v$ is a polygon (an S-node) then the embedding of $H_v$ is formed by orienting the graph $H_u$ for each
child node $u$ so that $p_u$ is placed on the outside of the polygon.
If $v$ is the root of the SPQR tree, this completes the proof that the embedding of $G$ is unique.
Otherwise, $s_vt_v$ must be on the outer face of $H_v$ (since it is an edge of the polygon.
Path $p_v$ must lie along the outer face of any embedding of $G$, because (if any child of $v$ is nontrivial) it contains vertices already required to lie along the outer face from lower levels of the SPQR tree.  If all children of $v$ are trivial, then $H_v$ is just the same polygon as $G_v$,
and separates two faces in any planar embedding of $G$; in this case
$p_v$ must lie along the outer face because it is not possible for two strictly convex internal faces to share a path of three or more vertices.

\item[Rigid case:]
In the final case, $G_v$ is a three-connected graph, which must be planar (else $G$ has no planar drawing).  Such graphs have a unique planar embedding up to the choice of outer face.
By the same reasoning as in the parallel case, each virtual edge must lie on the outer face,
or else it would be sandwiched between two internal faces leading to a nonconvexity in the drawing.
We divide into subcases according to the number of virtual edges.
\begin{itemize}
\item
If there are no virtual edges, then $G$ is itself 3-connected.  If $G$ is to be dual to a pseudoline arrangement with $L$ lines, then the outer face of $G$ must have $2L$ edges.  No other face of $G$ could have so many edges, because $G$ has at least four faces  and any internal face with $k$ edges would correspond to crossings between $(k/2)(k/2-1)/2$ pairs of pseudolines, leaving no crossings for the other faces.  So in this case the outer face can be uniquely identified as the face with the largest number of edges.  (In fact we can prove that no 3-connected graph has a symmetric planar drawing, but the proof is more complex, and we reuse this subcase's reasoning in the next subcase.)

\item
If there is a single virtual edge, it must be on the outer face, so this narrows down the choice of the outer face to two possibilities, the two faces of $G_v$ containing the virtual edge.  By the same reasoning as for the subcase with no virtual edges, these two faces must have differing numbers of edges and the outer face must be the one with the larger number of edges.  If $v$ is not the root, it has no children and $H_v=G_v$; otherwise, the embedding of $H_v$ is formed from that of $G_v$ by orienting the child of $v$ with $p_v$ along the outer face of $G_v$. 

\item
If there are two or more virtual edges, there can only be one face in $G_v$ containing these edges,
which must be the outer face of $G_v$.
The embedding of $H_v$ is fixed by placing the graph $H_u$ for each child $u$ of $v$ so that
the outer face of $H_u$ (minus the virtual edge connecting it to $G_v$) lies along the outer face of $G_v$.
\end{itemize}
\end{description}
\end{proof}

\begin{theorem}
Given an input graph $G$, we can determine whether $G$ is the dual of a weak pseudoline arrangement, and if so construct a symmetric planar drawing of $G$, in linear time.
\end{theorem}

\begin{proof}
If $G$ is biconnected, we choose a planar embedding of $G$ by Lemma~\ref{lem:2c-unique-embed}.
Otherwise, each articulation point of $G$ must be on the outer face of any embedding.
We find biconnected components of $G$, embed each component by Lemma~\ref{lem:2c-unique-embed}, and verify that these embeddings place the articulation points on the outer faces of each component.  We then connect the embeddings together into a single embedding having as its outer face the edges that are outer in each biconnected component; the choice of this embedding may not be unique but does not affect the correctness of our algorithm.

Finally, once we have an embedding of $G$, we must verify that we have the dual of a weak pseudoline arrangement (as the construction of Lemma~\ref{lem:2c-unique-embed} can also be applied to certain graphs that are not duals of weak pseudoline arrangements, such as odd polygons) and construct a symmetric planar drawing of it.  To do so, we first make sure all faces of $G$ are even.  We then apply the construction of Lemma~\ref{lem:spd2wpla}
(which needs only the embedding of $G$, not the actual drawing) to construct an arrangement of curves $\cal A$ dual to $G$ that should be a weak pseudoline arrangement.  We test that $\cal A$ has no closed curves, but not that it is an actual pseudoline arrangement.  We then apply the construction of Lemma~\ref{lem:wpla2spd}
to produce vertex placements for a drawing of $G$, test for each edge of $G$ that the endpoints of that edge are placed at unit distance apart with the expected slope, and test that each internal face of $G$ is drawn as a correctly oriented strictly convex polygon.  If $\cal A$ were not a weak pseudoline arrangement, either due to a curve self-crossing or to two curves crossing each other with the wrong orientation, this would result in the face of $G$ dual to that crossing point being drawn as a nonconvex polygon or an incorrectly oriented convex polygon, which we would detect, so if our input passes all these tests we have determined that it is the dual of a weak pseudoline arrangement and found a symmetric planar drawing.
\end{proof}

\begin{figure}[t]
\centering
\includegraphics[width=4.5in]{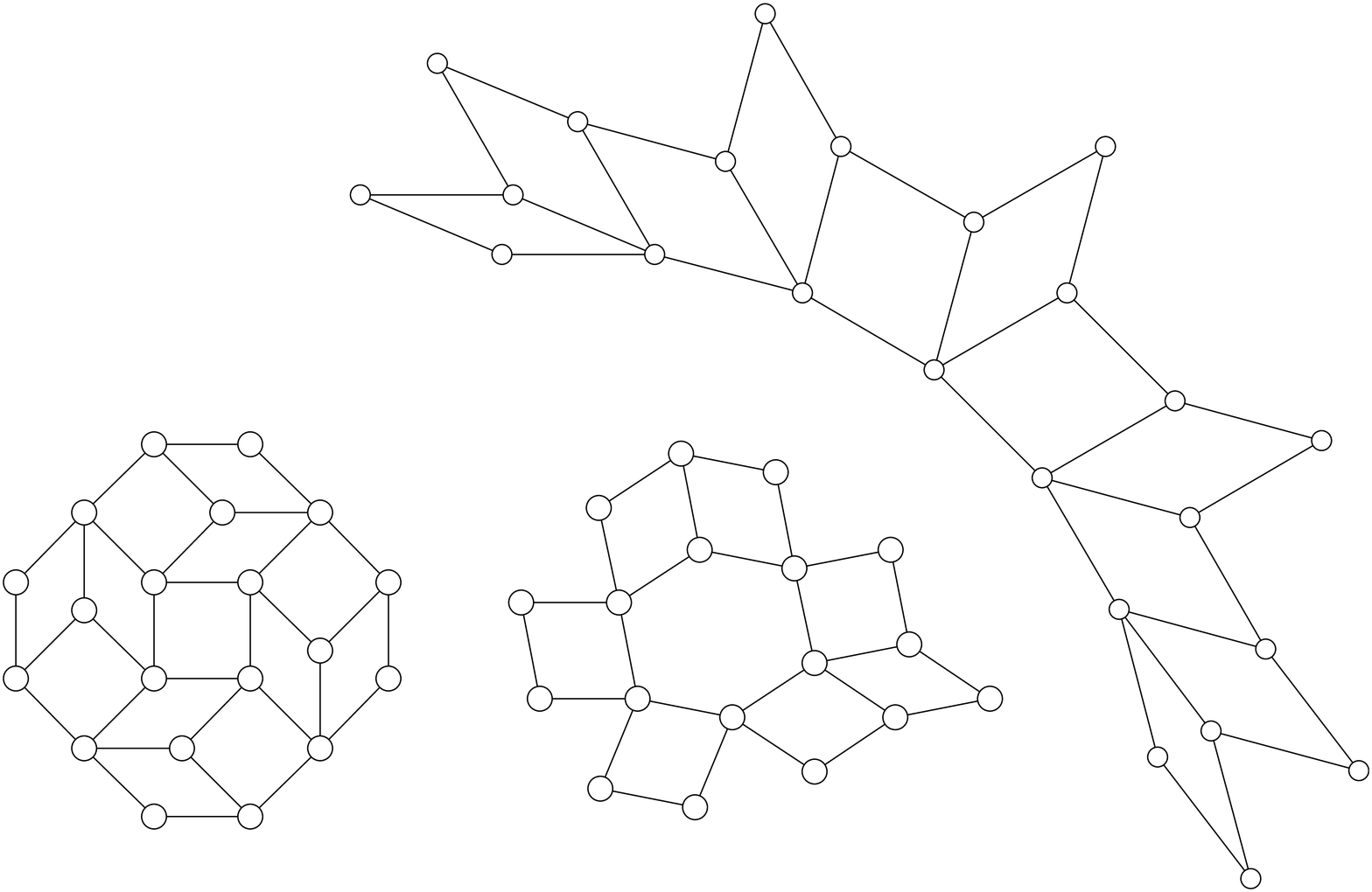}
\caption{Symmetric planar drawings of three irregular media.}
\label{fig:sfp}
\end{figure}

Our actual implementation is based on a simpler but less efficient algorithm that uses the known medium structure of the input to construct the dual weak pseudoline arrangement one curve at a time,
before applying the construction of Lemma~\ref{lem:wpla2spd} to produce a symmetric planar drawing from the weak pseudoline arrangement.
Examples of drawings produced by our symmetric planar drawing code
are shown in Figure~\ref{fig:sfp}.

\section{Implementation and examples}

\begin{figure}[t]
\centering
\includegraphics[width=4.5in]{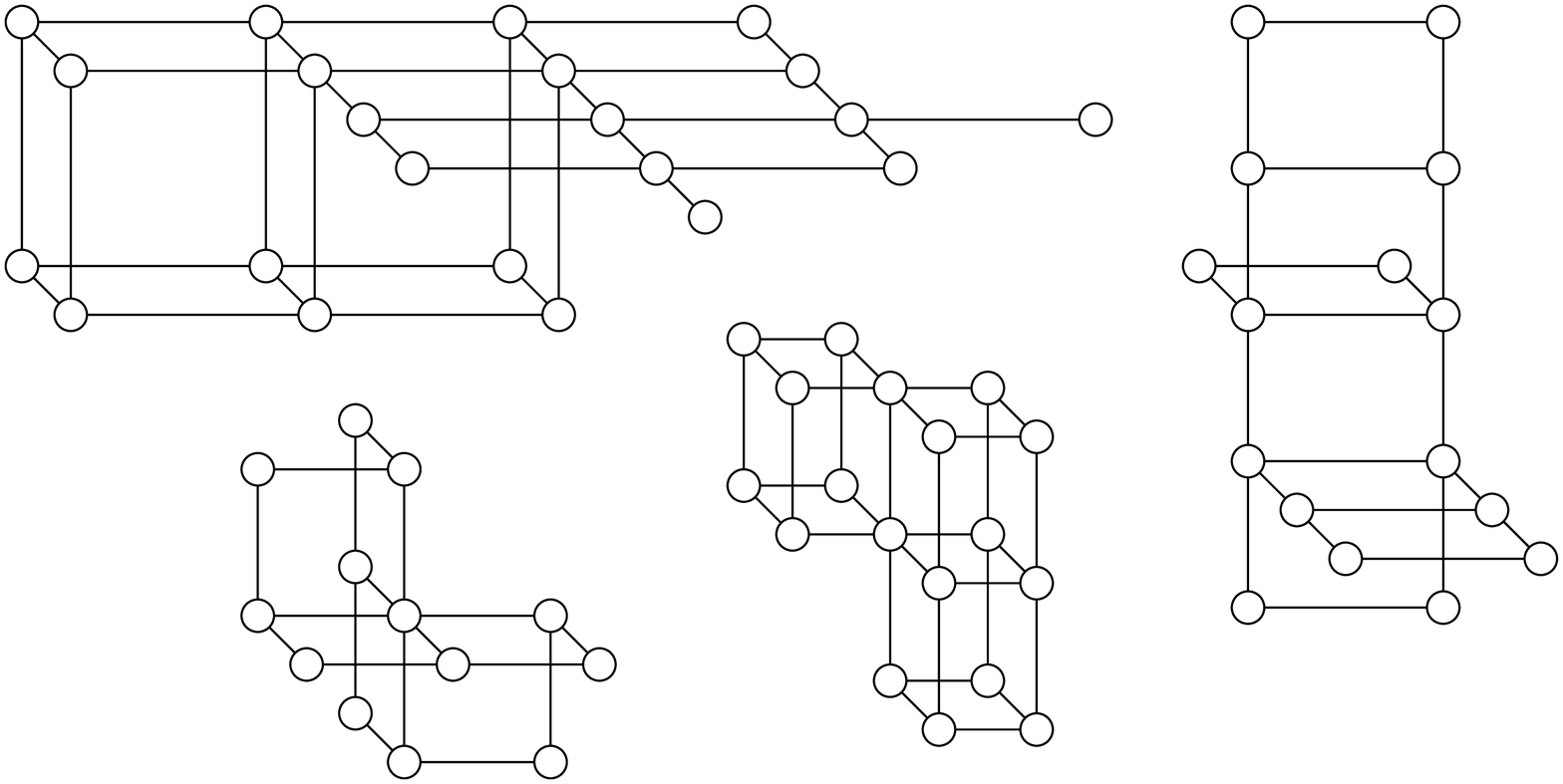}
\caption{Lattice drawings of four irregular media with three-dimensional lattice embeddings, from~\cite{Epp-EJC-?}.  The bottom left drawing is of a medium isomorphic to the weak ordering medium shown in Figure~\ref{fig:order}(right).}
\label{fig:grids3d}
\end{figure}

\begin{figure}[t]
\centering
\includegraphics[scale=0.175]{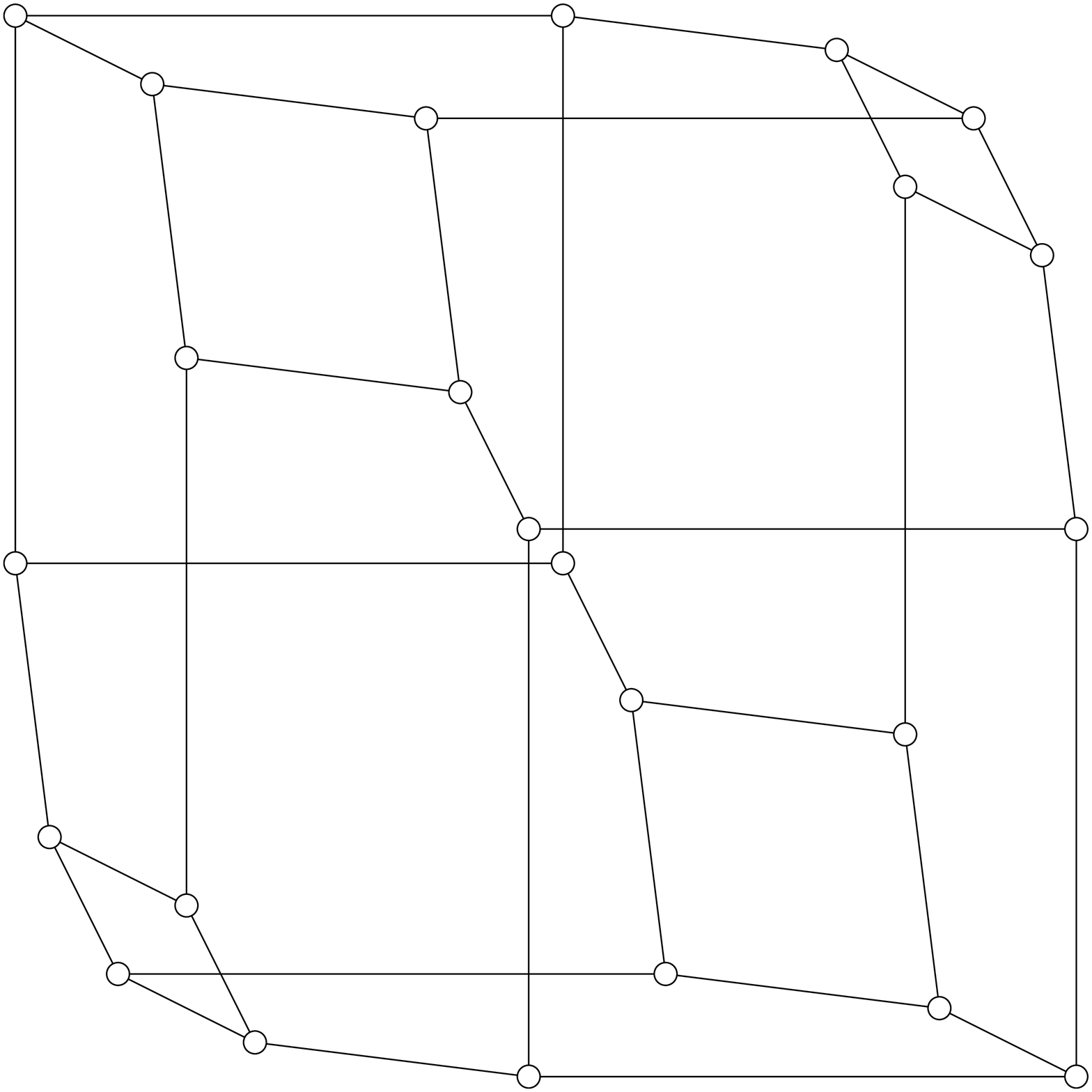}\qquad
\includegraphics[scale=0.7]{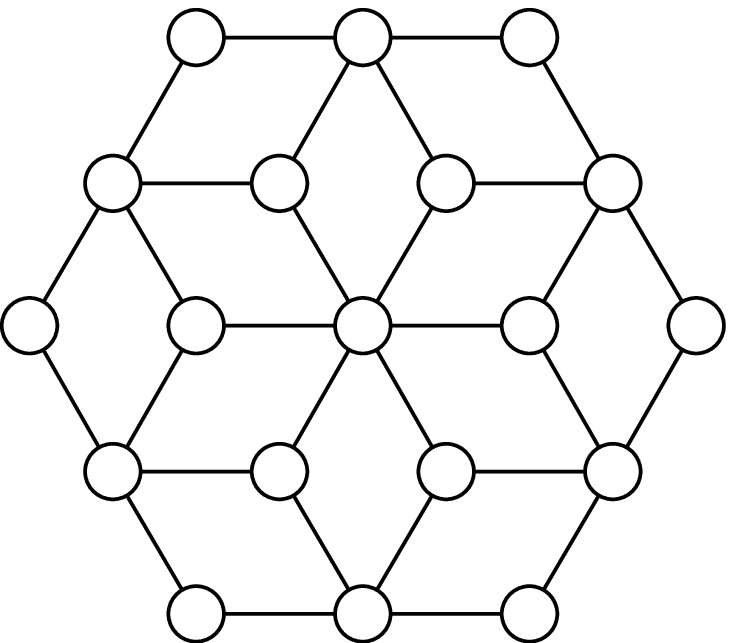}\qquad
\includegraphics[scale=0.7]{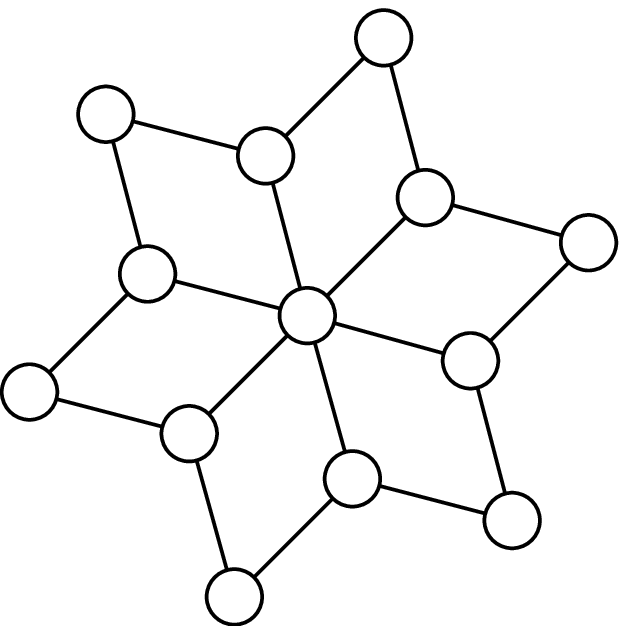}
\caption{Media defined by orderings of $n$-item sets. Left: Lattice drawing of total orderings (permutations) on four items. Center: Projected three-dimensional lattice drawing of partial orderings on three items.
Right: Symmetric planar drawing of weak orderings on three items.}
\label{fig:order}
\end{figure}

We implemented our algorithms in Python, with drawings output in SVG format.
Our code allows various standard combinatorial media (such as the collection of permutations on $n$ items) to be specified on the command line; irregular media may be loaded from a file
containing hypercube or lattice coordinates of each state.
We have seen already examples of our implementation's output
in Figures~\ref{fig:cube6}, \ref{fig:nosfp}, \ref{fig:sfp}, and~\ref{fig:grids3d}.
Figure~\ref{fig:order} provides additional examples.
All figures identified as output of our code have been left unretouched, with the exception that we have decolorized them for better printing.

\section{Conclusions and open problems}

We have shown several methods for drawing the state transition graphs of media.
There are several interesting directions future research in this area could take.

\begin{itemize}
\item If a three-dimensional lattice embedding has the property that it can be projected perpendicularly to the vector $(1,1,1)$ (or more generally $(\pm 1,\pm 1,\pm 1)$) without placing two vertices in the same point, we have seen (e.g. in Figure~\ref{fig:order}, center) that the projection produces a planar drawing with all edges having equal lengths and angles that are multiples of $60^\circ$.  Our lattice dimension algorithm can find a three-dimensional embedding, if one exists, and it is trivial to test the projection property.  However, a medium may have more than one three-dimensional embedding, some of which have the projection property and some of which don't.  For instance, the medium in the lower right of Figure~\ref{fig:grids3d} is the same weak ordering medium as the one in Figure~\ref{fig:order}(right),
however the former drawing is from a lattice embedding without the projection property while the latter drawing could be a projection of a different lattice embedding (although it was actually produced by our symmetric planar drawing algorithm).  Is it possible to efficiently find a projectable three-dimensional lattice embedding, when one exists?  More generally, given an arbitrary dimension lattice embedding of a medium, can we find a planar projection when one exists? 

\item We have seen that hypercubes may be drawn in $O(n^2)$ area such that vertices and nonadjacent edges have unit separation, and all edges coming from actions of the same token are drawn as line segments that are translates of each other.  Is it possible to achieve the same translate property and $O(n^2)$ area bound for more general classes of media?

\item Our lattice and symmetric planar drawings have several desirable qualities; for instance, all edges corresponding to a single token are drawn as line segments with the same slope and length,
and our lattice drawings have good vertex-vertex and vertex-edge separation.
However, we have not seriously examined the extent to which other important graph
drawing properties may be achieved.  For instance,
$d$-dimensional hypercubes (and therefore also media with up to $2d$ tokens)
may be drawn with geometric thickness~\cite{Epp-thickness} at most $\lceil d/3\rceil$ (Figure~\ref{fig:cube6}, right)
however our lattice projection methods achieve geometric thickness only $\lceil d/2\rceil$ while the only way we know how to achieve the better $\lceil d/3\rceil$ bound is to use a more irregular drawing
in which edges coming from the same token are no longer parallel.
\end{itemize}

\raggedright
\bibliographystyle{abuser}
\bibliography{media}

\end{document}